\documentclass[12pt]{article}
\usepackage{amssymb}
\usepackage{amsmath}

\begin{document}


\title{Sum rules for odd and even states of confining potentials}
\author{C.V.Sukumar\\
Department of Physics, University of Oxford\\
Thoretical Physics, 1 Keble Road, Oxford OX1 3NP}
\maketitle

\begin{abstract}
Using the Green's function associated with the one-dimensional
Schr{\"{o}}dinger equation it is possible to establish a hierarchy of sum
rules involving the eigenvalues of confining potentials which have
only a bound state spectrum. For some potentials the sum rules could lead to
divergences. It is shown that when this happens it is possible to
examine the separate sum rules satisfied by the even and odd
eigenstates of a symmetric confining potential and by subtraction cancel the divergences exactly and produce
a new sum rule which is free of divergences. The procedure is illustrated by
considering symmetric power law potentials and the use of several examples. One of the examples considered shows that
the zeros of the Airy function and its derivative obey a sum rule and
this sum rule is verified. 
\end{abstract}

\noindent

\vfill\eject

\noindent
\section{\noindent\textbf{Introduction}}
Green's functions $G$ are used to study the solutions to inhomogeneous
differential equations which satisfy homogeneous boundary conditions.
$G$ satisfies the same boundary conditions as the solution $\Psi$ of a
second order differential equation. G
satisfies a corresponding differential equation with a delta function
source term (Morse and Feshbach (1953)). For one-dimensional second order differential equations there are two
methods for constructing the Green's function. In the first method $G$
is constructed as a sum involving the eigenfunctions and eigenvalues of
the homogeneous differential equation. In the second method $G$ is
constructed from two solutions of the homogeneous differential equation
satisfying appropriate boundary conditions. The existence of two
alternative ways of constructing the Green's function leads to the
establishment of certain equalities involving the eigenfunctions and
eigenvalues of the differential operator. For studying the bound states of
quarks with definite angular momenta the radial Schr{\"{o}}dinger equation
with a confining potential has been used extensively. Such solutions
for zero orbital angular momentum ({\it viz}) the s-states, can be viewed
as the odd states of a symmetric confining potential in the infinite
space $-\infty \le x \le \infty$ (Quigg {\it et al} (1980)). Since confining potentials have no
scattering states and have only bound states the corresponding Green's
functions lead to a hierarchy of sum rules which only involve the
bound state eigenvalues. Such sum rules can be powerful tools to
establish the extent to which  a set of numerically computed eigenvalues
approach
completeness. A leading order sum rule  has been used recently by Mezincescu (2000) to
shed light on the hypothesis advanced by Bender and Boettcher (2001) that a cubic
oscillator with imaginary coupling strength has a completely real
spectrum.

The sum rules involving all the bound state eigenvalues do not always lead to convergent expressions  and prove useful. When this happens it is still possible to extract something useful from the Green's functions. In this paper it is shown that it is possible to establish separate sum rules for the odd and even states  in a symmetric potential by considering Green's functions satisfying different boundary conditions and that when divergences in the sums arise they may be of the same type for the odd and even states. If this is the case then it is possible to subtract the two divergent sum rules to establish a convergent sum rule involving both the odd and even states. The plan of this paper is as follows: the construction of the Green's function and the derivation of the sum rules for the inverses of the eigenvalues of confining potentials are presented in Section 2. The sum rules for the even and odd states in symmetric confining potentials of the power law form are constructed in Section 3. The spectral zeta function for power law potentials has been studied by Voros (2000). In Section 4 several solvable examples are considered and the sum rules are verified. In Section 5 the case of an arbitrary symmetric confining potential is considered and general results are presented.

\noindent
\section{\protect\bigskip  Green's Functions and Sum Rules}
Consider a second order differential operator of the form
\begin{equation}
L\left(x\right) = - \frac{d^{2}}{dx^{2}} + V\left( x \right) \label{1}
\end{equation}
with a complete set of solutions $\Psi_{n}$ of the eigenvalue equation
\begin{equation}
L\Psi_{n} = \lambda_{n} \Psi_{n}  \label{2}
\end{equation}
for a choice of two boundary conditions $B_{1}$ and $B_{2}$ on $\Psi$ at
$x_{1}$ and $x_{2}$ respectively. $B_{1}$ and $B_{2}$ can be chosen to
be homogeneous boundary conditions of the Neumann or Dirichlet kind,
i.e., the eigenfunction or its derivative vanishes at $x_{1}$ or
$x_{2}$. The orthonormality of the eigenfunctions $\Psi_{n}$ is given by
\begin{equation}
\int_{x_{1}}^{x_{2}} \Psi_{m}\left(x\right) \Psi_{n}\left(x\right)
=\delta_{mn} . \label{3}
\end{equation}
A Green's function $G$ satisfying the same boundary conditions $B_{1}$
and $B_{2}$ as $\Psi_{n}$ at $x_{1}$ and $x_{2}$ , respectively, may be
defined as a solution of
\begin{equation}
\left(L\left(x\right) - \lambda \right) G\left(x,y;\lambda\right) =
\delta\left(x-y\right)  .\label{4}
\end{equation}
A representation of $G$ in terms of the complete set of eigenfunctions
and eigenvalues is discussed in standard textbooks on Green's
functions. For a general $V\left(x\right)$ the complete set of
eigenfunctions include eigenstates for both discrete and continuous
eigenvalues. However, if $V\left(x\right)$ is a confining potential,
then the complete set is made up of only discrete eigenvalues
$\lambda_{n}$ and G may be represented as a discrete sum given by
\begin{equation}
G\left(x,y;\lambda\right) =\sum_{n}
\frac{\Psi_{n}\left(x\right)\Psi_{n}\left(y\right)}{\lambda_{n} -
\lambda} . \label{5}
\end{equation}
Using the completeness relation
\begin{equation}
\sum_{n}\Psi_{n}\left(x\right) \Psi_{n}\left(y\right) = \delta\left(x
- y \right) \label{6}
\end{equation}
and eq.(2) it is easy to verify that $G$ defined by eq.(5) satisfies
eq.(4).

For one-dimensional equations such as eq.(4) an alternative procedure to
construct $G$ is to solve eq.(4) directly. Let $\phi_{1}$ and $\phi_{2}$
be two linearly independent solutions of 
\begin{equation}
\left( L - \lambda\right) \phi_{1,2} = 0     \label{7}
\end{equation}
chosen such that $\phi_{1}$ satisfies the boundary condition $B_{1}$ but
not $B_{2}$ and $\phi_{2}$ satisfies $B_{2}$ but not $B_{1}$. Then
in terms of $\phi_{1}$ and $\phi_{2}$, $G$ may be constructed as
\begin{equation}
G\left(x,y;\lambda\right) = -\frac{\phi_{1}\left(x_{<},\lambda\right)\phi_{2}\left(x_{>},\lambda\right)}{W\left(\phi_{1},\phi_{2}\right)}  \label{8}
\end{equation}
where the Wronskian $W$ given by
\begin{equation}
W\left(\phi_{1},\phi_{2}\right) =
\phi_{1}\left(x,\lambda\right)\frac{d\phi_{2}\left(x,\lambda\right)}{dx}
- \phi_{2}\left(x,\lambda\right)
\frac{d\phi_{1}\left(x,\lambda\right)}{dx} \label {9}
\end{equation}
satisfies 
\begin{equation}
\frac{dW}{dx} = 0  \label{10}
\end{equation}
and $x_{<} \left(x_{>}\right)$ is the smaller (larger) of (x,y). Using
eqs.(7) and (10) it is easy to verify that $G$ defined by eq.(8) is a
solution of eq.(4). When the Green's function can be constructed from
eq.(8) then eq.(8) may be taken to be the definition of $G$ and eq.(5)
may be viewed as the expression of an equality relating $G$ to the
eigenfunctions and eigenvalues. The existence of two expressions for $G$
leads to certain equalities. Using the orthonormality condition defined
by eq.(3) it can be shown from eq.(5) that
\begin{equation}
\int_{x_{1}}^{x_{2}} G\left(x,x;\lambda\right) dx  = \sum_{n}
\frac{1}{\left(\lambda_{n} - \lambda\right)} . \label{11}
\end{equation}
Eq.(11) is an expression of a sum rule if eq.(8) for $G$ is used to
evaluate the integral. The above procedure can be generalized to provide
a hierarchy of sum rules as discussed by Sukumar (1990). The
orthonormality of the eigenfunctions can be shown to lead to a second
order sum rule in the form
\begin{equation}
\int_{x_{1}}^{x_{2}} dx  \int_{x_{1}}^{x_{2}} dy
\ G\left(x,y;\lambda\right)G\left(y,x;\lambda\right)
=\sum_{n}\frac{1}{\left(\lambda_{n}-\lambda\right)^{2}}. \label{12}
\end{equation}
Similar sum rules involving a chain of $n$ Green's functions starting
from $x$ and returning to $x$ via ($n-1$) intermediate set of points
may be used to establish sum rules of arbitrary order $n$. The sums over
the powers of the inverses of the eigenvalues
for the case $\lambda =0$ are referred to as generalized zeta functions
in the literature.  All the sum rules defined so far converge for finite
values of $x_{1}$ and $x_{2}$. For general confining potentials with
$x_{1}\to -\infty$ and $x_{2}\to \infty$ extra convergence conditions
are needed to ensure that the sum rules lead to convergent expressions. These questions are examined in the next section.
\noindent
\section{\protect\bigskip Sum Rules for Power Law Potentials}
Consider symmetric potentials of the power law form ${\vert x \vert}^{N}$.
The eigenstates split into groups of odd states satisfying the boundary
conditions
\begin{equation}
{Lt}_{x\to 0}\Psi\left(x\right) \ \to \ 0\  , {Lt}_{x\to\infty} \Psi\left(x\right)
\to \ 0
\label{13}
\end{equation}
and even states satisfying the boundary conditions
\begin{equation}
{Lt}_{x\to 0}\frac{d }{dx} \Psi\left(x\right)  \to \ 0\ ,\  {Lt}_{x\to\infty}
\Psi\left(x\right) \ \to \ 0\ . \label{14}
\end{equation}
If the parameter $\lambda$ is chosen to be  equal to $0$ then the zero
energy Schr{\"{o}}dinger equation 
\begin{equation}
\frac{d ^{2}}{dx^{2}} \Psi = {\vert x \vert}^{N} \Psi \label{15}
\end{equation}
under the substitutions
\begin{equation}
\nu = \frac{2}{N+2} \ , \ z = \nu x^{\frac{1}{\nu}} \ , \  \beta
=\frac{1}{N+2}\ ,\  \Psi= z^{\beta} \Phi   \label{16}
\end{equation}
transforms into 
\begin{equation}
\left(\frac{d^{2}}{dz^{2}}  + \frac{1}{z} \frac{d }{dz}  -
\left(1+{\frac{\beta^{2}}{z^{2}}} \right) \right)\ \Phi =0 \label{17}
\end{equation}
which is the differential equation satisfied by the modified Bessel
functions ( Abramowitz and Stegun (1965)). Hence the solutions are of the
form
\begin{equation}
\Psi\  = {\sqrt x} \ I_{\pm\beta}\left(\nu x^{\frac{1}{\nu}}\right)
\label{18}
\end{equation}
from which we can construct solutions which satisfy appropriate boundary
conditions as $x\to0$ and $x\to\infty$. The solution which shows
decaying behaviour as $x\to \infty$ is ${\sqrt x}
K_{\beta}\left(z\right)$ , the solution which vanishes as $x\to 0$ is
${\sqrt x} I_{+\beta}\left(z\right)$ and the solution for which the
derivative vanishes as $x\to 0$ is ${\sqrt x}
I_{-\beta}\left(z\right)$. Hence the Green's function appropriate for
the description of eigenstates satisfying boundary conditions analogous
to those in eq. (13) is
\begin{equation}
G_{1}\left(x,y\right)\ = \ \nu {\sqrt x}{\sqrt y} I_{+\beta}\left(\nu
x_{<}^{\frac {1}{\nu}}\right) \ K_{\beta}\left(\nu x_{>}^{\frac {1}{\nu}}\right)
\label{19}
\end{equation}
and the Green's function appropriate for the description of eigenstates
satisfying boundary conditions analogous to those in eq.(14) is
\begin{equation}
G_{2}\left(x,y\right)\ =\ \nu {\sqrt x}{\sqrt y} I_{-\beta}\left(\nu
x_{<}^{\frac {1}{\nu}}\right)\ K_{\beta}\left(\nu x_{>}^{\frac {1}{\nu}}\right)\ .
\label{20}
\end{equation}
The boundary conditions satisfied by $G_{1}$ and $G_{2}$ are
\begin{align}
{Lt}_{x\to0}\ G_{1}\left(x,y\right)\ \to \ 0 \ ,\ \ {Lt}_{x\to\infty}\
G_{1}\left(x,y\right)\ \to \ 0 \notag\\
{Lt}_{x\to0}\ \frac{d }{dx} G_{2}\left(x,y\right)\ \to\ 0\ ,\ 
{Lt}_{x\to\infty}\ G_{2}\left(x,y\right)\ \to \ 0 \ . \label{21}
\end{align}
The WKB result for the eigenvalues $\lambda_{n}$ in the limit
of large quantum number $n$ can be derived from the Bohr-Sommerfield
formula and is of the form
\begin{equation}
\lambda_{n}\ =\ \left(\left(n+\frac{1}{2}\right)\frac{{\sqrt{\pi}}\left(N+2\right)\Gamma\left(\frac{N+2}{2N}\right)}{2\Gamma\left(\frac{1}{N}\right)}\right)^{\frac{2N}{N+2}}
\label{22}
\end{equation}
which shows that the sums over the inverses of the eigenvalues will not
converge if $N<2$. This also indicates that the integrals of the Green's
functions over the infinite domain $\left[0,\infty\right]$ will not
converge if $N<2$. When the Green's functions in eqs.(19) and (20) are used in eq.(11) with the choice of limits $x_{1}=0$ and $x_{2}=\infty$ the resulting integrals
can be performed (Gradshteyn and Ryzhik (1965)) if $N>2$ ({\it i.e})
$\beta<\frac{1}{4}$. The resulting sum rules are
\begin{equation}
S_{1}\ =\ \sum_{n=0}^{\infty}\frac{1}{\lambda_{2n+1}}\ =\
\beta^{2-4\beta}\
\frac{\Gamma\left(3\beta\right)\Gamma\left(2\beta\right)\Gamma\left(1-4\beta\right)}{\Gamma\left(1-2\beta\right)\Gamma\left(1-\beta\right)} \ \ ,\  \ 
\beta<\frac{1}{4} \label{23}
\end{equation}
for the odd states and
\begin{equation}
S_{2}\ =\ \sum_{n=0}^{\infty}\frac{1}{\lambda_{2n}}\ =\
\beta^{2-4\beta}\ \frac{\Gamma\left(2\beta\right)\Gamma\left(\beta\right)\Gamma\left(1-4\beta\right)}{\Gamma\left(1-3\beta\right)\Gamma\left(1-2\beta\right)}
\ \ ,\  \ \beta<\frac{1}{4} \label{24}
\end{equation}
for the even states. The WKB approximation for the eigenvalues given by
eq.(22) indicates that when $N\le 2$ the sums given by $S_{1}$ and
$S_{2}$ diverge but it also indicates that if we look at the difference
$S_{2}-S_{1}$ the terms in the resultant sum have the large $n$ behaviour of
$n^{-\frac{3N+2}{N+2}}$ which would lead to convergent sums for all
positive definite values of $N>0$. Hence we examine
\begin{equation}
S_{2}\ -\ S_{1}\ =\ \int_{0}^{\infty} \
\left(G_{2}\left(x,x\right)\ -\ G_{1}\left(x,x\right)\right) dx  \ .
\label{25}
\end{equation}
Using the addition formulae satisfied by the modified Bessel functions which
appear in the integrand the integral can be
written as
\begin{equation}
S\ =\ S_{2}\ -\ S_{1}\ =\ \frac{4\beta}{\pi}\ \sin \pi\beta\
\int_{0}^{\infty}x\ K_{\beta}^{2}\left(\nu x^{\frac{1}{\nu}}\right)\ dx
\label{26}
\end{equation}
and the resulting integral can be performed (Gradshteyn and Ryzhik (1965)) to
obtain the result
\begin{equation}
S\ =\ \sum_{n=0}^{n=\infty}\
\frac{\left(-\right)^{n}}{\lambda_{n}}\ =\ \beta^{2-4\beta}\
\frac{\Gamma\left(3\beta\right){\Gamma^{2}\left(2\beta\right)}\Gamma\left(\beta\right)}{\Gamma\left(4\beta\right)}\ \frac{\sin\pi\beta}{\pi} \
.\label{27}
\end{equation}
By using the reflection property satisfied by the gamma functions 
\begin{equation}
\Gamma\left(z\right)\ \Gamma\left(1-z\right)\ =\ \frac{\pi}{\sin\pi z}
\label{28}
\end{equation}
the results for the sum rules can be simplified. $S$ can be represented
in the form
\begin{equation}
S\ =\ \sum_{n=0}^{\infty}\ \frac{\left(-\right)^{n}}{\lambda_{n}}\ =\ \beta^{2-4\beta}\
\frac{\Gamma\left(3\beta\right)\Gamma^{2}\left(2\beta\right)}{\Gamma\left(4\beta\right)\Gamma\left(1-\beta\right)} \ ,\ N>0\ ,\ \beta<\frac{1}{2} .\label{29}
\end{equation}
In terms of $S$ the other sum rules derived in this section may be written in the compact form
\begin{align}
S_{2}\ &=\ \sum_{n=0}^{\infty}\ \frac{1}{\lambda_{2n}}\ \ \ =\ S\ \frac{\sin
3\pi\beta}{\sin \pi\beta}\ \frac{1}{2\cos 2\pi\beta} \ ,\ \  N>2\ \ ,\
\ \ \left(\beta<\frac{1}{4}\right),  \\
S_{1}\ &=\ \sum_{n=0}^{\infty}\ \frac{1}{\lambda_{2n+1}}\ =\ S\
\frac{1}{2\cos 2\pi\beta} \ ,\ \  N>2\ \ ,\ \left(\beta<\frac{1}{4}\right)\ .
\label{}
\end{align}
We have thus provided a direct proof of the sum rules listed by Voros (2000) by a method that is capable of being generalized to other symmetric confining potentials.  In the next section special values of the power law index $N$ for which the
eigenvalues are readily available in the literature are considered and
the resulting sum rules are examined.

\noindent
\section{\protect\bigskip Examples of Sum Rules}
\subsection{Linear Potential V(x)=x}
For the linear potential the Schr{\"{o}}dinger equation for
zero energy is given by eq.(15) with $N=1$ and can be identified as the
differential equation satisfied by Airy functions (Abramowitz and
Stegun (1965)). The eigenvalues of the even eigenstates whose derivatives vanish at $x=0$ are
given by the negative of the zeros of the derivative of the Airy function and the 
eigenvalues of the odd eigenstates whose wavefunctions vanish at $x=0$
are given by the negative of the zeros of the Airy function. The zeros of the Airy
function and its derivative are listed in Abramowitz and Stegun (1965) and the
first 10 zeros are given below in Table 1.

\medskip
\begin{table}[h]
\begin{center}
\begin{tabular}{|l|l|l|l|}
\hline
$n$ & $\lambda_{2n}$ & $\lambda_{2n+1}$ & $\lambda_{2n}^{-1}\
-\ \lambda_{2n+1}^{-1}\smallskip$ \\ \hline
0 &  1.01879 &  2.33811 &  0.55386 \\ \hline
1 &  3.24820 &  4.08795 &  0.06324 \\ \hline
2 &  4.82010 &  5.52056 &  0.02632 \\ \hline
3 &  6.16331 &  6.78671 &  0.01490 \\ \hline
4 &  7.37218 &  7.94413 &  0.00977 \\ \hline
5 &  8.48849 &  9.02265 &  0.00697 \\ \hline
6 &  9.53545 & 10.04017 &  0.00527 \\ \hline
7 & 10.52766 & 11.00852 &  0.00415 \\ \hline
8 & 11.47506 & 11.93602 &  0.00337 \\ \hline
9 & 12.38479 & 12.82878 &  0.00279 \\ \hline
\end{tabular}
\end{center}
\caption{The eigenvalues of the lowest 10 even and odd eigenstates for
the linear potential and the differences between the inverses of the
eigenvalues taken in pairs}
\end{table}

The eigenvalues of the states with larger values of $n$ can be accurately
estimated from the asymptotic formulae
\begin{align}
Lt_{n\to\infty}\ \lambda_{2n}\ &=\ \left(\frac{3\pi
n}{2}\right)^{\frac{2}{3}}\ \left(1\ +\
\frac{1}{4n}\right)^{\frac{2}{3}} \\
Lt_{n\to\infty}\ \lambda_{2n+1}\ &=\ \left(\frac{3\pi
n}{2}\right)^{\frac{2}{3}}\ \left(1\ +\
\frac{3}{4n}\right)^{\frac{2}{3}}        \label{}
\end{align}
which show that the sums $S_{1}$ and $S_{2}$ diverge. However if we look
at the difference $S$ between the sums over the inverses of the even and
odd 
eigenvalues the terms in $S$ have the asymptotic behaviour
\begin{equation}
Lt_{n\to \infty}\ \left(\ \lambda_{2n}^{-1}\ -\ \lambda_{2n+1}^{-1}\
\right)\ =\ \left(\frac{2}{3\pi n}\right)^{\frac{2}{3}}\ \frac{1}{3n}
\label{}
\end{equation}
which indicates that the integral over n would converge. The first
10 terms in the series for $S$ are tabulated in the last column of Table
1 and add up to $\sim$0.691. The remaining terms in the sum can be added by
converting the sum over $n$ into an integral over $n$ and using the
asymptotic estimate given by eq.(34) for the integrand. Such an evaluation gives
\begin{equation}
R\ =\ \int_{n}^{\infty} \left(\ \lambda_{2n}^{-1}\ -\ \lambda_{2n+1}^{-1}\
\right)\ dn\ =\ \frac{1}{2}\ \left(\frac{2}{3\pi
n}\right)^{\frac{2}{3}} \ .\label{}
\end{equation}
Hence an estimate of $S$ is given by choosing $n=10.5$ in eq.(35)
which gives the result that
\begin{equation}
S\ = \sum_{n=0}^{\infty}\ \left(\ \lambda_{2n}^{-1}\ -\
\lambda_{2n+1}^{-1}\ \right)\ \sim \ .691 \ +\ 0.037\ =\ 0.728
\label{}
\end{equation}
in good agreement with the exact result derivable from eq.(29) 
using the value $\beta=\frac{1}{3}$ for $N=1$ which gives
\begin{equation}
S\ =\ \left(\frac{1}{3}\right)^{\frac{2}{3}}\
\frac{\Gamma\left(\frac{2}{3}\right)}{\Gamma\left(\frac{4}{3}\right)}\
\sim\ 0.729\ .\label{}
\end{equation}
By calculating more zeros of the Airy function and its derivative exactly
the sum involving the eigenvalues can be evaluated more accurately
without approximating part of the sum by an integral over $n$ and the
agreement between the two sides of eq.(29) can be verified to more
significant figures.

The asymptotic limits given by eqs.(32) and (33) also indicate that
even though the sums over the inverses of the eigenvalues will not
converge they also show that the sums over squares and higher powers of
the inverse of the eigenvalues will converge. Hence higher order sum
rules such as those arising from eq.(12) may be verified by using
the appropriate eigenvalues in the sum and the appropriate Green's functions in the
integral.
\noindent

\subsection{Simple Harmonic Oscillator  V(x)\ =\ $x^{2}\ +\ 1$ }
For the oscillator potential $V\left(x\right)=x^{2}$ the eigenvalues are
given by $\lambda_{n}=2n+1$. The sums $S_{1}$ and $S_{2}$ are both
divergent but the difference $S$ gives
\begin{equation}
S\ =\ 1\ -\ \frac{1}{3}\ +\ \frac{1}{5}\ -\ \frac{1}{7}\ +\ ....\ =\
\frac{\pi}{4}   \label{}
\end{equation}
in agreement with the value from eq.(29) for $\beta=\frac{1}{4}$
corresponding to $N=2$. The zero energy
solutions to the Schr{\"{o}}dinger equation have to be calculated numerically to verify the integral
involving the Green's functions. However analytical results are
possible for the oscillator shifted in energy by one unit. Therefore we
consider this case in detail. For the shifted simple harmonic oscillator potential $V\left(x\right)=x^{2}+1$ the eigenvalues
are $\lambda_{n}=2n+2$ with $n$ taking positive even and odd values. The
sums $S_{2}$ and $S_{1}$ over the inverses of the even and odd eigenvalues are both
logarithmically divergent. However the difference between the sums gives
\begin{equation}
S\ =\ \sum_{n=0}^{\infty}\ \frac{{\left(-\right)}^{n}}{\lambda_{n}}\ =\
\frac{1}{2}\ -\ \frac{1}{4}\ +\ \frac{1}{6}\ -\ \frac{1}{8}\ +....\ =\
\frac{\ln2}{2}\ . \label{}
\end{equation}
The zero energy solutions to the Schr{\"{o}}dinger
equation for this potential are gaussians and integrals over gaussians 
and they may be used to construct the Green's
functions
\begin{align}
G_{1}\left(x,x\right)\ &=\ -\frac{2}{\sqrt\pi}\ e^{x^{2}}\
\int_{0}^{x}e^{-y^{2}}\ dy \ \int_{\infty}^{x}e^{-z^{2}}\ dz \\
G_{2}\left(x,x\right)\ &=\ -e^{x^{2}}\ \int_{\infty}^{x}\ e^{-y^{2}}\ dy
\\
G_{2}\left(x,x\right)\ -\ G_{1}\left(x,x\right)\ &=\ \frac{2}{\sqrt\pi}\ e^
{x^{2}} \
\int_{\infty}^{x}e^{-y^{2}}\ dy\ \int_{\infty}^{x} e^{-z^{2}}\ dz \ .
\label{38}
\end{align}
The sum rule $S$ in this case gives rise to the relation
\begin{equation}
{\sqrt\pi}\ \int_{0}^{\infty}\ e^{x^{2}}\ \left(\ 1\ -\
erf\left(x\right)\ \right)^{2}\ dx\ =\ {\ln2}  \label{}
\end{equation}
where erf(x) is the error function defined by
\begin{equation}
erf\left(x\right)\ =\ \frac{2}{\sqrt\pi}\ \int_{0}^{x}\ e^{-y^{2}}\ dy\
. \label{40}
\end{equation}
This is an interesting integral relation which has arisen from the sum
rule for the shifted oscillator potential.

\noindent

\subsection{Quartic Potential V(x)\ =\ $x^{4}$}
The low lying eigenvalues for the even states in the quartic potential
can be obtained from Bender ${\it et\ al}$ (1977) and the eigenvalues for
the odd states can be extracted from Hioe and Montroll (1975) by
appropriate renormalization of the eigenvalues to accommodate the changed
strength of the potential. The lowest eigenvalues are listed in
Table 2.
\medskip

\begin{table}[h]
\begin{center}
\begin{tabular}{|l|l|l|}
\hline
$n$ & $\lambda_{2n}$ & $\lambda_{2n+1}$ \\ \hline 
0 &  1.060362 &  3.799673 \\ \hline
1 &  7.455698 & 11.644746 \\ \hline
2 & 16.261826 & 21.238373 \\ \hline
3 & 26.528472 & 32.098598 \\ \hline
4 & 37.923001 & 43.981158 \\ \hline
\end{tabular}
\end{center}
\caption{The low lying eigenvalues for the quartic potential}
\end{table}

The eigenvalues for the higher lying states may be calculated using the
WKB formula in eq.(22) for $N=4$ to give
\begin{align}
\lambda_{2n}\ &\sim\ \alpha \ \left(n\ +\ \frac{1}{4}
\right)^{\frac{4}{3}} \\
\lambda_{2n+1}\ &\sim \ \alpha \left(n\ +\
\frac{3}{4}\right)^{\frac{4}{3}}  \label{}
\end {align}
where
\begin{equation}
\alpha\ =\ \left(6\sqrt\pi\
\frac{\Gamma\left(\frac{3}{4}\right)}{\Gamma\left(\frac{1}{4}\right)}\right)^{\frac{4}{3}} \ . \label{}
\end{equation}
To verify the sum rules the first $k$ terms may be summed exactly using
the numerical estimates in Table 2 and the remainder may be evaluated by
approximating by integrals over $n$ and using the WKB estimates which
give 
\begin{align}
\int_{k+\frac{1}{2}}^{\infty}\ \lambda_{2n}^{-1}\ dn\ &=\
\frac{3}{\alpha}\ \left(k\ +\ 
\frac{3}{4}\right)^{-\frac{1}{3}}  \\
\int_{k+\frac{1}{2}}^{\infty}\ \lambda_{2n+1}^{-1}\ dn\ &=\
\frac{3}{\alpha}\ \left(k\ +\ \frac{5}{4}\right)^{-\frac{1}{3}} \ .
\label{}
\end{align}
Setting $k=4$ we get
\begin{align}
S_{1}\ &\sim\ 0.45003\ +\ 0.31349\ &=\ 0.76352  \\
S_{2}\ &\sim\ 1.22266\ +\ 0.30413\ &=\ 1.52679  \\
S    \ &\sim\ 0.77263\ -\ 0.00936\ &=\ 0.76327 \label{}
\end{align}
to be compared with the values from eqs.(29)-(31) calculated using
$\beta=\frac{1}{6}$ corresponding to $N=4$ which give
\begin{equation}
S_{1}\ =\ S\ =\ \frac{S_{2}}{2}\ =\ \left(\frac{1}{6}\right)^{\frac{4}{3}}\
\frac{\Gamma\left(\frac{1}{2}\right)\Gamma^{2}\left(\frac{1}{3}\right)}{\Gamma\left(\frac{2}{3}\right)\Gamma\left(\frac{5}{6}\right)}\ =\ 0.76330
\label{}
\end{equation}
in good agreement with the numerical results. The agreement can be
improved by calculating more of the eigenvalues exactly. 

\noindent

\subsection{Particle In a Box  $N\to\infty$}

To check the sum rules for the case $N\to\infty$ we proceed as follows.
The sum rules scale when the strength of the power law potential is changed by a
factor $\gamma$. The scaling factor for the energy eigenvalue is easily found to be
$\gamma^{\frac{2}{N+2}}$. Therefore if we consider the potential
\begin{align}
Lt_{N\to\infty}\ \left(\frac{2|x|}{\pi}\right)^{N}\ &=\ 0\ \ if\ \ x\ <\ \frac{\pi}{2} \notag  \\
                 &=\ 1\ \ if\ \ x\ =\ \frac{\pi}{2}  \notag \\
                 &=\ \infty\ \ if\ \ x\ >\ \frac{\pi}{2} \label{}
\end{align}
which corresponds to a particle in a box with infinite potential at the
walls at $|x|=\frac{\pi}{2}$. The eigenvalues for this potential are the
squares of integers. Hence the sums over the inverses of the eigenvalues
can all be found in closed form and are given by
\begin{align}
S_{1}\ &=\ \sum_{n=0}^{\infty} \left(2n+2\right)^{-2}\ = \
\frac{\pi^{2}}{24} \notag \\
S_{2}\ &=\ \sum_{n=0}^{\infty} \left(2n+1\right)^{-2}\ =\
\frac{\pi^{2}}{8} \notag \\
S   \ &=\ S_{2}\ -\ S_{1}\ =\ \frac{\pi^{2}}{12} \ . \label{}
\end{align}
When the scaling factor is taken into account the sum rule in eq.(29)
becomes
\begin{equation}
S\ =\ {Lt}_{\beta\to 0} \ \left(\frac{\pi\beta}{2}\right)^{2-4\beta}\
\frac{\Gamma\left(3\beta\right){\Gamma}^{2}\left(2\beta\right)}{\Gamma\left(4\beta\right)\Gamma\left(1-\beta\right)} \ \ . \label{}
\end{equation}
If the limiting behaviour of the gamma function $Lt_{z\to 0} \sim
z^{-1}$ is used then eqs.(56),(30) and (31) give
\begin{equation}
S\ =\ 2S_{1}\ =\ \frac{2}{3}S_{2}\ =\ \frac{\pi^{2}}{12}  \label{}
\end{equation}
in agreement with the results in eq.(55).
\noindent

\section{Conclusions}

In this paper it has been shown that for symmetric confining potentials it is
possible to establish sum rules involving the eigenvalues of
odd and even states separately. It has been shown that even when the sum
rules for even and odd states do not converge the difference between
them could converge. We have shown that for power law potentials simple
expressions for the sum rules may be derived which show an interesting
structure. Even though the eigenvalues for $N\ne2$ have to be found
numerically the sum rules show that the inverses of the eigenvalues add
up to some simple ratios involving the well known gamma functions. This
is an interesting result from a mathematical point of view. Higher order
sum rules such as those emerging from the application of eq.(12) indicate
even more intricate relationships between the solutions of the
Schr{\"{o}}dinger equation.

\bigskip
\vfill\eject

\section{References} 

\noindent[1] Morse, P. \& Feshbach, H. 1953 {\it Methods of Theoretical
Physics} (New York: McGraw-Hill) Vol 1, 791-811.

\noindent[2] Quigg, C., Thacker, H.B. \& Rossner, J.L. 1980 
 {\it Phys. Rev.} {\bf D 21}, 234-240.

\noindent[3] Mezincescu, G.A. 2000 {\it J.Phys.A} {\bf 33} 4911-6.

\noindent[4] Bender, C.M. and Boettcher, S. 1998 
 {\it Phys. Rev. Let.} {\bf 80} 5243-6.

\noindent[5] Voros, A. 2000 {\it J.Phys. A} {\bf 33} 7423-50.

\noindent[6] Sukumar, C.V. 1990 {\it Am. J. Phys.} {\bf 58}, 561-5.

\noindent[7] Abramowitz, M. \& Stegun, I.A.\ 1965 {\it Handbook of Mathematical Functions} (New York: Dover) 437-78.

\noindent[8] Gradshteyn, I.S. \& Ryzhik, I.M. 1965 {\it Table of Integrals,
Series and Products}, (New York: Academic) 693-4.

\noindent[9] Bender, C.M., Olaussen, K. \& Wang, P.S. 1977 
 {\it Phys. Rev.} {\bf D 16}  1740-8.

\noindent[10] Hioe, F.T. \& Montroll, E.W. 1975 
 {\it J. Math. Phys.} {\bf 16} 1945-55.

\end{document}